\documentclass[pss]{wiley2sp} % provides pss two-column style
\usepackage{amsmath}
\titlefigure[scale=0.3]{GammaMod}

\newcommand{\erf}{\frac{\text{erf}\left(\omega r_{12}\right)}{r_{12}}}

\newcommand{\dr}{\,d\vec{ r}}
\newcommand{\dro}{\,d\vec{r}_1}

\newcommand{\drodrt}{\,d\vec{r}_1d\vec{r}_2}
\newcommand{\br}{\vec{ r}}

\newcommand{\bo}{\vec{r}_1}

\newcommand{\bt}{\vec{r}_2}

\newcommand{\bth}{\vec{r}_3}
\newcommand{\bfo}{\vec{r}_4}

\begin{document}

% Title of the article
\title{Range separated functionals in the density functional based
  tight binding method: Formalism}

% Abbreviated title for the page headers
\titlerunning{Range separated functionals in DFTB }

% Authors
\author{%
  Thomas A. Niehaus\textsuperscript{\Ast,\textsf{\bfseries 1}} and Fabio Della Sala \textsuperscript{2}
}

% Abbreviated list of authors for the page headers
\authorrunning{Thomas A. Niehaus}

%E-mail-address of corresponding author
\mail{e-mail
  \textsf{thomas.niehaus@physik.uni-regensburg.de}}

% author's affiliations/addresses
\institute{%
  \textsuperscript{1}\,Department of Theoretical Physics, University of Regensburg, 93040 Regensburg, Germany \\ 
   \textsuperscript{2}\, Istituto Nanoscienze-CNR, Via per Arnesano, 73100, Lecce \& Center for Biomolecular Nanotechnologies @UNILE, 
Istituto Italiano di Tecnologia (IIT), Via Barsanti, 73010 Arnesano  (LE), Italy \\
}

\received{XXXX, revised XXXX, accepted XXXX} % do not change, will be filled in by the publisher
\published{XXXX} % do not7 change, will be filled in by the publisher

% Please select about four verbal keywords for your manuscript.

\abstract{%
% This is a macro for the typesetting of two-column text in an
% abstract. It will typeset the two arguments in \abstcol{}{} as the
% left and right column inside the abstract box. At the
% columnbreak there will be always a columnbreak (\par), so both
% columns start with a new paragraph. No automatic column height
% balancing is done.
%
% If used with a \titlefigure it will silently output both
% parameters as consecutive paragraphs.
%
% The macro is defined exclusively inside the argument of \abstract{};
% if used outside it will raise an error.
%
% Usage: \abstcol{<left column>}{<right column>}

%  \titlefigure{fig/GammaMod}
%   }
\abstcol{%
  A generalization of the density-functional based tight-binding method (DFTB) for the 
  use with range-separated exchange-correlation functionals
  is presented. It is based on the Generalized Kohn-Sham (GKS) formalism and  
  employs the density matrix as basic variable in the expansion of the
  energy functional, in contrast to the traditional DFTB scheme. 
  The GKS-TB equations are derived and appropriate integral
  approximations are discussed in detail.  
  Implementation issues and numerical aspects of the new scheme are also covered. 
  }
{%}
}
}

% The class file requires the standard graphicx Latex package. See the 'LaTeX
% standard graphics and color packages documentation' for more information at
% <http://tug.ctan.org/tex-archive/macros/latex/required/graphics/grfguide.pdf>.
%
% Accepted figure file formats depend on which LaTeX flavour is used.
% Classic LaTeX is always able to use Encapsulated Postscript (EPS);
% PDFLaTeX can't use this but accepts PDF, JPG, PNG, and GIF formats.
%
% See examples for implementing graphics in floating figure environments later in this file.
% If \titlefigure is given, it takes as its mandatory parameter the
% name (without extension) of some figure file.

\maketitle   % please do not remove
\section{Introduction}
Over the last two decades the density functional based tight-binding (DFTB) method found widespread use 
in such different areas as computational chemistry, condensed matter physics, biophysics and the 
ever growing field of nanoscience. DFTB is an approximate density functional theory (DFT) that is characterized 
by a simplified energy functional and additional integral approximations. These modifications 
give rise to a highly reduced computational cost maintaining at the same time a useful accuracy 
for many applications. Starting with the work of Seifert \cite{Seifert1986}, the original DFTB 
formalism has been generalized in multiple directions.  The group of Thomas Frauenheim has been
 particularly active in this respect. Examples are the self-consistent extension 
of DFTB \cite{elstner1998scc}, the treatment of spin-polarized systems \cite{Kohler2001,Kohler2007} a
nd van der Waals interactions \cite{Elstner2001}, the combination 
with the non-equilibriums Greens function theory in quantum transport \cite{DiCarlo2005}, or
 the extension to time dependent DFT(B) \cite{Niehaus2001a,Niehaus2005,Niehaus2009} and the 
GW formalism of many body perturbation theory \cite{Niehaus2005a}.

A general feature of DFTB is that it often parallels the accuracy of DFT for different problem classes, 
inheriting also spectacular failures of the latter. A large number of these problematic cases 
can be traced back to the self-interaction error (SIE) of popular local or semi-local exchange-correlation (xc) 
functionals\cite{sciencesic}. In Hartree-Fock (HF) theory the interaction of an electron with itself is exactly canceled 
by opposite terms in the exchange part of the Fock matrix. As exchange is approximated in 
the local density or gradient corrected approximations for the xc potential, there is a  residual 
self-interaction. This leads to wrong asymptotics of the Kohn-Sham potential and 
an overly broadened density \cite{almbladh85,prlasy}. Signatures of this deficiency are seen in such different areas 
as the incorrect dissociation of radical cations \cite{Chai2008}, 
the instability  of polaronic defects \cite{Niehaus2004}, 
the incorrect description of organic-metal interfaces \cite{neaton06,fabiano09}
or the underestimation of charge  transfer excited states \cite{dreuw2004ftd,Wanko2004}. 
In the field of molecular electronics the
self-interaction error has also been found to contribute strongly to the 
significant overestimation of conductances \cite{sanvito05,Ke2007}.

In recent years so called range-separated or long-range corrected \cite{gill96,Savin1996,savin97,handy04,hirao01,Baer2010} xc functionals have been found 
to alleviate the above mentioned difficulties. Going back to Gill \cite{gill96} and Savin \cite{Savin1996}, the idea is 
to split the electron-electron interaction into short-range (sr) and long-range (lr) contributions
 \begin{equation}
  \label{sepr12}
  \frac{1}{r_{12}} = \frac{1-\text{erf}(\omega r_{12})}{r_{12}} + \frac{\text{erf}(\omega r_{12})}{r_{12}},
\end{equation}
where the long range part is treated exactly, while the short range
part gives rise to a modified pure density functional. In
Eq.~(\ref{sepr12}), the term ``$\text{erf}$'' denotes the error function
and $\omega$ is an empirical parameter. The resulting theory may be viewed as a generalization 
of the well known hybrid functionals \cite{B3} with fixed weighting coefficients of density functional and 
HF exchange. 
 In this way, the error compensation of density functionals for exchange and 
correlation is kept in the short range, whereas the self-interaction error is removed at least asymptotically 
through the long range contribution.

Here it should be noted that in the range-separated functionals optimized for solid-state systems, 
it is often the short-range part that is treated exactly, whereas the long-range
part is treated as a density-functional. Such methods allow for an accurate
description of fundamental band gaps
\cite{Heyd2003,Heyd2006,Henderson2010} and avoid the known artefacts
of HF exchange for metallic solids \cite{Gerber2007}. Since we are mostly interested in
the improvement of DFTB for finite nanostructures and molecules, we
 keep with the separation of Eq. \ref{sepr12} in
the following.     
%metals 
%or small gap materials 
%the long range exchange is often modified or completely abandoned.} 
Note also that both range-separated and hybrid methods belong to the
Generalized Kohn-Sham (GKS) \cite{seidlgks} scheme, as both contain a non-local potential  
(i.e. a fraction of the non-local HF exchange operator)  in the single particle equations.

Several applications and modification of the original subdivision in Eq. \ref{sepr12} 
 have emerged in the past years and show that optimized range separated functionals not only alleviate 
the above-mentioned problems but also compete with the best hybrid functionals in terms of standard applications 
like structure prediction and thermochemistry \cite{Chai2008,Vydrov2006}. Moreover it must be underlined
that only functionals which contains a full HF contribution in the long-range can correctly describe 
charge-transfer excited states \cite{tawada2004lrc,baer09,wong09}.

The goal of this investigation is to extend the theoretical foundations of the DFTB scheme, currently limited to local or semi-local xc-functionals,
to range-separated functionals.
%because the theory is currently adapted to local density functionals only. 
The corresponding modified energy functional is presented in section (\ref{efunc}), 
while in section (\ref{gks}) effective GKS equations are derived. 
Sections (\ref{happrox}) and  (\ref{reprep}) deal with additional approximations 
for the derived terms in the spirit of the traditional DFTB scheme. The paper closes with a brief summary and outlook.

\def\slatWF{\Phi}
\def\ext{{\text{ext}}}
\def\vext{v^\ext}

\def\tloc{\text{loc}}
\def\vloc{v^\tloc}

\def\txc{\text{xc}}
\def\tx{\text{x}}
\def\ttot{\text{tot}}
\def\th{\text{H}}
\def\tnn{\text{NN}}
\def\txlr{\text{xlr}}
\def\vxlr{v^\txlr}

\def\txsrc{\text{xsr+c}}
\def\vxsrc{v^\txsrc}
\def\fxsrc{f^\txsrc}

\section{Total energy expression}
\label{efunc}
In the GKS \cite{seidlgks} formalism the total energy $E_\ttot$ can be rewritten 
as a functional of the density matrix 
\begin{equation}
  \label{ddm}
  \gamma(\bo,\bt) = 2\sum_i^{N/2} \psi_i(\bo) \psi_i^*(\bt),
\end{equation}
which equals the electron density $\rho(\br)$ on the diagonal, i.e.
$\rho(\br) = \gamma(\br,\br)$.
The wave functions $ \psi_i$ denote the spatial part of GKS spin-orbitals 
and we confine the treatment to the special case
of closed shell systems with $N$ electrons
and $N/2$ occupied orbitals.
We thus have: 
\begin{equation}
\label{etot}
E_\ttot[\gamma]=T[\gamma]+E_\txc[\gamma]+E_\th[\rho]+E_\ext[\rho]+E_\tnn
\end{equation}
%where the We start the discussion with the definition of the spinless Dirac
%density matrix for a system of N electrons:
where
\begin{eqnarray}
T[\gamma]&=&  \int \left.\left( -\frac{1}{2} \vec{\nabla}_{r_2}^2 \right) 
\gamma(\bo,\bt)\right|_{\bt = \bo} \dro \\
E_\th[\rho] &=& \frac{1}{2} \iint \frac{\rho(\bo) \rho(\bt)}{r_{12}} \drodrt  \\
E_\ext[\rho] &=& \int \rho(\br) \vext(\br) \dr \\ 
E_\tnn &=& \frac{1}{2}\sum_{AB}\frac{ Z_A Z_B}{\left| \vec{R}_A - \vec{R}_B \right| }
\end{eqnarray}
represent the kinetic, Hartree, external and nuclear-repulsion energy, respectively.
% With Eq.~(\ref{ddm}), the total
%energy may be written as:
%
%\begin{eqnarray}
%\label{etot}
%E_\text{tot} &=&  \int \left.\left( -\frac{1}{2} \vec{\nabla}_{r_2}^2 + v^\text{ext}
%  (\bt)\right) \rho(\bo,\bt)\right|_{\bt = \bo} \dro \nonumber \\
%&+&  E^H[\rho] +
%E^\text{xc}[\rho]  + E^\text{ii},
%\end{eqnarray}
%
%with the external potential $ v^\text{ext}(\br)$, the ion-ion repulsion
%\begin{equation}
%  E^\text{ii} = \frac{1}{2}\sum_{AB}\frac{ Z_A Z_B}{\left| \vec{R}_A -
%      \vec{R}_B \right| },
%\end{equation}
%
%the Hartree energy ($\vec{r}_{12} = \bo - \bt$)
%\begin{equation}
%  E^H[\rho] = \frac{1}{2} \iint \frac{\rho(\bo) \rho(\bt)}{r_{12}} \drodrt,
%\end{equation}
%and the exchange-correlation energy $E^\text{xc}[\rho]$.
In the range-separated formalism the exchange  is divided into a short-range
(xsr) and long-range (xlr) contribution and the xc energy functional reads:
\begin{equation}
 E_\txc[\gamma] =  E_\text{xsr}[\rho] +E_\txlr[\gamma]+ E_\text{c}[\rho] 
\end{equation}
where the xlr exchange is given explicitly as a functional of the density matrix
\begin{equation}
  \label{exlr}
  E_\txlr[\gamma] = -\frac{1}{4} \iint \erf \gamma(\bo,\bt) \gamma(\bt,\bo) \drodrt.
\end{equation}

Note that in the GKS formalism the auxiliary system of partially-interacting electron 
is still described by a single-Slater determinant
\cite{seidlgks}, thus only the exchange is range-separated. The whole electron-electron interaction is
range-separated in the the multi-determinant extension of the Kohn-Sham theory \cite{tolouse04,tolouse05}.
%also $E^\text{c}$ should be range
%separated which allows to introduce hybrid methods that combine DFT
%with correlated wave function based methods\cite{tolouse04} without the risk of
%over-counting correlation. 
As we are aiming at an approximate scheme
that is foremost self-interaction free, it is sufficient to remain in the GKS scheme
%this route is not followed
%here and we stay with the standard correlation density functionals  in
and to employ local or gradient corrected approximations for the correlation.
%full interaction
%range. 
Explicit functional forms of the $\omega$-dependent short-range
exchange for the LDA and the
Perdew-Burke-Ernzerhof density functionals are available in the
literature \cite{hirao01,Toulouse2004,Henderson2008}. 
Please note that the
special form of the interaction kernel $\text{erf}(\omega r)/r$ in Eq.~(\ref{exlr}) is not
without alternative. Besides the error function also Yukawa
and other screened potentials have been used to define interaction
ranges. It turns out that the precise choice of the separation
function is of minor importance \cite{Baer2010}. 
From a numerical
point of view,   the form of
Eq.~(\ref{exlr}) is advantageous both for basis functions of Gaussian type
or calculations using plane waves, since it
allows for an efficient evaluation of the required two-electron integrals.

As the first approximation we now expand the total energy of
Eq.~(\ref{etot}) around a certain reference density matrix
$\gamma_0(\bo,\bt)$ up to second order in $\Delta\gamma =\gamma-\gamma_0$:
\begin{equation}
\label{exp}
E_\ttot =  E^\text{(0)} + E^\text{(1)} + E^\text{(2)} +{\cal
  O}(\Delta\gamma^3),
\end{equation}
which parallels the expansion in terms of the density in the
conventional DFTB scheme. The individual terms in Eq.~(\ref{exp}) are
given as:
\begin{eqnarray}
 E^\text{(0)} &=&
T[\gamma_0]  
+ E_\ext[\rho_0]
%   \int \left.\left( -\frac{1}{2} \vec{\nabla}^2_{r_2} + v^\text{ext}
%  (\bt)\right) \rho_0(\bo,\bt)\right|_{\bt = \bo} \dro \nonumber \\
% &+& 
+ E_\th[\rho_0] 
+ E_\txc[\gamma_0]  + E_\tnn \label{E0} \\
%%%%%%%%%%%%%%%%%%%%%%%%%%%%%%%%%%%%%%%%%%%%%%%%%%%%%%%%%%%%%%%%%%%%%%%%%%%%
E^\text{(1)} &=&  \int \left.\left( -\frac{1}{2} \vec{\nabla}^2_{{\bf r}_2} 
  (\bt)\right) \Delta\gamma(\bo,\bt)\right|_{\bt = \bo} \dro \nonumber \\
&+& \int \vext(\br)  \Delta\rho(\br) \dr \nonumber \\
&+&  \iint \frac{\rho_0(\bt)
  \Delta\rho(\bo)}{r_{12}} \drodrt \nonumber\\
&+& \iint \left.\frac{ \delta E_\txc[\gamma]}{\delta \gamma(\bo,\bt)}
\right|_{\gamma=\gamma_0} \Delta\gamma(\bo,\bt) \drodrt \\
%%%%%%%%%%%%%%%%%%%%%%%%%%%%%%%%%%%%%%%%%%%%%%%%%%%%%%%%%%%%%%%%%%%%%%%%%%%%
E^\text{(2)} &=&
  E_\th[\Delta\rho]
+ \frac{1}{2}\iiiint \left.\frac{ \delta^2 E_\txc[\gamma]}{\delta \gamma(\bo,\bt)\delta \gamma(\bth,\bfo)}
\right|_{\gamma=\gamma_0} \nonumber\ \\
&\times& \,\Delta\gamma(\bo,\bt) \Delta\gamma(\bth,\bfo) \,d\bo
d\bt d\bth d\bfo.
\end{eqnarray}

These formulas may be simplified by introducing potentials $v^\alpha$
and kernels $f^\alpha, \alpha\in\{ \text{xsr; xlr; c} \}$ as first and
second order functional derivatives of the xc-energy with respect to the
density matrix, respectively. 
The terms with $\alpha\in\{ \text{xsr; c} \}$ are both local can be treated together:
%locality of the functionals leads to the expressions
\begin{eqnarray}
 \frac{ \delta E_\txsrc[\gamma] }{\delta \gamma(\bo,\bt)} &=&
  \delta(\vec{r}_{12}) \vxsrc[\rho](\bo) \\
%\end{equation}
%and 
%\begin{equation}
\frac{ \delta^2 E_\txsrc[\gamma]}{\delta \gamma(\bo,\bt)\delta \gamma(\bth,\bfo)} &=&
\delta(\vec{r}_{12}) \delta(\vec{r}_{34})  \fxsrc[\rho](\bo)\delta(\vec{r}_{13}), \nonumber \\
\end{eqnarray}

while the non-local long-range exchange potential and kernels take the form:
\begin{eqnarray}
  \frac{ \delta E_\txlr[\gamma]}{\delta \gamma(\bo,\bt)}&=& \vxlr[\gamma](\bo,\bt) \\
                                                          &=&    -\frac{1}{2}\erf \gamma(\bt,\bo) \\
  \frac{ \delta^2 E_\txlr[\gamma]}{\delta \gamma(\bo,\bt)\delta \gamma(\bth,\bfo)} 
%  \frac{ \delta E_\txlr[\gamma]}{\delta \gamma(\bo,\bt)}&=& v_\txlr[\gamma](\bo,\bt)  
&=&  -\delta(\vec{r}_{12}) \delta(\vec{r}_{34}) \frac{1}{2}\erf
\end{eqnarray}

Defining also the local part of the GKS potential as
\begin{equation}
\vloc[\rho](\br) = \vext(\br)+ v^\th[\rho](\br)+  \vxsrc[\rho](\br)
\end{equation}
%$ + v_\tee$, with $v^\text{ee}$ given as
%\end{equation}
%\begin{eqnarray}
%  v^\text{ee}[\rho](\bo,\bt) &=& \left\{ v^H[\rho](\bo) +
%    v^\text{xsr+c}[\rho](\bo) \right\} \delta(\vec{r}_{12})
% \nonumber\\ &+& v^\text{xlr}[\rho](\bo,\bt),
%\end{eqnarray}
%and $ v^\text{xsr+c} =  v^\text{xsr} +  v^\text{c}$, we arrive at
the first and the second order term of the energy becomes

\begin{eqnarray}
E^\text{(1)} &=& \int \left.\left( -\frac{1}{2} \vec{\nabla}^2_{{\bf r}_2} 
  (\bt)\right) \Delta\gamma(\bo,\bt)\right|_{\bt = \bo} \dro \nonumber \\
  \label{E1}
 &+&  \int   \vloc[\rho_0](\br) \Delta\rho(\br) \dr  \\
 &+&  \iint  \vxlr[\gamma_0](\bo,\bt) \Delta\gamma(\bo,\bt) \drodrt  \nonumber 
\end{eqnarray}
and
\begin{eqnarray}\label{E2}
E^\text{(2)} &=&  \frac{1}{2}\iint \left(\frac{1}{r_{12}} +
  \fxsrc[\rho_0](\bo)\delta(\vec{r}_{12}) \right) \nonumber \\
&&\times 
\Delta\rho(\bo) \Delta\rho(\bt) \drodrt  \\
&-& \frac{1}{4}\iint \erf  \Delta\gamma(\bo,\bt)\Delta\gamma(\bt,\bo) \drodrt.\nonumber
\end{eqnarray}
%the zero order contribution to the energy $E^\text{(0)}$ being
%unchanged with respect to Eq.~(\ref{E0}).
In the next section we will derive tight-binding approximations to Eqs. (\ref{E1}) and (\ref{E2}).

\section{Generalized Kohn-Sham Tight-Binding}
\label{gks}
Similar to the conventional DFTB approach, the reference density matrix $\gamma_0$ for the molecular system of interest 
is obtained by a superposition of atomic quantities. To this end a full DFT calculation in the 
range-separated formalism (RS-DFT) is performed for neutral spin-unpolarized atoms.
\begin{equation}
\label{atom}
\left[ \hat{t} + \hat{v}^\tloc + \hat{v}^\txlr + \hat{v}^\text{con}\right] |\phi_\mu\rangle = \epsilon_\mu |\phi_\mu\rangle.
\end{equation}
The atomic orbitals $\phi_\mu$ are then used as basis functions for
the molecular problem in a linear combination of atomic orbitals
(LCAO) ansatz for the desired GKS molecular orbitals:
\begin{equation}
|\psi_i\rangle = \sum_{\mu} c_{\mu i} |\phi_\mu\rangle .
\end{equation}
 Usually only the
valence orbitals are included in the expansion and each atomic orbital
is given as a converged superposition of Slater type orbitals. The
additional confining potential $\hat{v}^\text{con}= (r/r_0)^2$ in
Eq.~(\ref{atom}) has been found to provide improved basis sets in the
traditional DFTB approach \cite{porezag1995ctb,Frauenheim2002}. 
In order to derive the GKSTB equations in this basis, we introduce the corresponding AO density matrix $P_{\nu\mu}$:
 \begin{equation}
  \gamma(\bo,\bt) = \sum_{\nu\mu} P_{\nu\mu}  \phi_\nu(\bo) \phi_\mu(\bt),
\end{equation}
with analogous definitions holding for $\gamma_0$ and $\Delta\gamma$. Here $\mu = \{Alm\}$ is a compound index indicating
the atom on which the basis function is centered, its
angular momentum $l$ and magnetic quantum number $m$. The reference density matrix has the simple form
$
P_{\nu\mu}^0  = \delta_{\nu\mu} n_\mu,
$
where $n_\mu$ denotes the occupation of orbital $\phi_\mu$ in the atom. Using the expansion of the Kohn-Sham states 
 with yet to be determined molecular orbital (MO) coefficients $c_{\mu i}$, we also have the relation
 \begin{equation}
   P_{\nu\mu} = 2  \sum_{i}^{N/2} c_{\nu i} c^*_{\mu i}.
\end{equation}

The total energy of Eqs.~(\ref{E0}, \ref{E1}, \ref{E2}) depends on the MO coefficients through the density matrix as given in the previous equation. Variation with respect to $c_{\mu i}^*$ under the constraint of orthonormality of the KS states leads to the following generalized eigenvalue problem:
\begin{equation}
  \sum_\nu H_{\mu\nu}c_{\nu i} =  \epsilon_i \sum_\nu S_{\mu\nu}c_{\nu i}.
\end{equation}

Here $S_{\mu\nu}$ denotes the overlap between two basis functions and matrix elements of the effective Hamiltonian are explicitly given as:
\begin{equation}
  H_{\mu\nu} =  H_{\mu\nu}^\text{(0)} + H_{\mu\nu}^\text{(1)}
\end{equation}

with the zero order contribution
\begin{eqnarray}
\label{hmn0}
 H_{\mu\nu}^\text{(0)} = \int \phi_\mu^*(\br) \left(  -\frac{1}{2}
    \vec{\nabla}^2 + \vloc[\rho_0](\br)\right)
  \phi_\nu(\br) \dr \nonumber  \\
%+  \int  \phi_\mu(\br)   v_\text{loc}[\rho_0](\br)  \phi_\nu(\br) \dr \\
+  \iint  \phi_\mu(\bo) \Big(  \vxlr[\gamma_0](\bo,\bt)  \Big ) \phi_\nu(\bt) \drodrt, 
\end{eqnarray}

and a first order potential shift
that has full range (fr) and long range (lr) components

\begin{eqnarray}
 H_{\mu\nu}^\text{(1)} &=&  \Delta v^\text{fr}_{\mu\nu} + \Delta v^\text{lr}_{\mu\nu} \nonumber \\
&=&  \iint \phi_\mu^*(\bt) \left(\frac{1}{r_{12}} +
 \fxsrc[\rho_0](\bo)\delta(\vec{r}_{12})
%  f^\text{xsr+c}[\rho_0](\bo,\bt)  
\right) \Delta\rho(\bt)  \nonumber \\
&& \times 
\phi_\nu(\bo) \drodrt\label{hmn1} \\
&-& \frac{1}{2}  \iint \phi_\mu^*(\bt) 
\Big ( \vxlr[\Delta\gamma](\bo,\bt)
%  \erf \Delta\gamma(\bt,\bo)
 \Big )
\phi_\nu(\bo) 
\drodrt.\nonumber
\end{eqnarray}
The matrix elements in Eq.~(\ref{hmn0}) depend only on the known reference density matrix $\gamma_0$, while 
the terms in Eq.~(\ref{hmn1}) stem from the second order term in the total energy and 
feature the difference density matrix $\Delta\rho$. Because of this, the eigenvalue problem 
has to be solved self-consistently as in a regular DFT calculation. In the following sections 
further approximations to $H_{\mu\nu}^\text{(0)}$ and $H_{\mu\nu}^\text{(1)}$  are proposed 
that make the method numerically more efficient.
\section{Approximations for the Hamilton matrix elements}
\label{happrox}
\subsection{The zero order Hamiltonian $H_{\mu\nu}^\text{(0)}$}
In the spirit of the traditional DFTB scheme we evaluate $H_{\mu\nu}^\text{(0)}$ of Eq.~(\ref{hmn0}) in the following two-center approximation:
\begin{eqnarray}
\label{DFTBme}
H_{\mu \nu}^\text{(0)} =
\left\{
\begin{array}{c@{\quad :\quad}l}
 \epsilon_{\mu}^{\text{free atom}} & \mu = \nu \\
 \langle\phi_{\mu} | \hat{t}
 +
  \hat{v}^\tloc_{AB}+ \hat{v}^\txlr_{AB} |\phi_{\nu}
  \rangle  &\mu\, \epsilon\, A, \nu\, \epsilon\, B \\
 0 & \text{otherwise}
\end{array}
\right.
\end{eqnarray}
where $\hat{v}^\tloc_{AB}$ and $\hat{v}^\txlr_{AB}$ are evaluated from a reference density matrix 
given by the sum of the density matrices of atom A and B. 
Both crystal field effects (the change of the on-site matrix elements due to the potential of neighboring atoms) 
and three center integrals are neglected in  Eq.~(\ref{DFTBme}). 
As the potentials are decaying much faster in the traditional DFTB (where the GGA xc-potential is exponentially decaying), 
these approximations are certainly more critical in the present approach and possibly need further consideration. 
Please note also that the zero order terms alone do not provide the correct asymptotics. 
The potential in Eq.~(\ref{DFTBme}) decays as $-1/r-1/r=-2/r$ (from the sum of two isolated atoms) 
instead of the desired $-1/r$ behavior. 
%The correct potential is restored by inclusion of the second order contribution to the Hamiltonian, 
%since already at this level the expansion of the Hartree and exchange energy is exact.

Like in empirical tight-binding schemes, the zero order matrix
elements of DFTB are precomputed and stored in Slater-Koster tables
\cite{Slater1954} as a function of distance between atoms A and
B. This is convenient since many matrix elements vanish by symmetry
(the integrals are non-zero only if $\phi_\mu$ and $\phi_\nu$ share
the same magnetic quantum number) and the rotation to the molecular
frame can be accomplished by simple transformation rules. For the
range-separated formalism discussed here the availability of such a
tabulation is not obvious as the potentials are non-local. However, it
can be shown that the Slater-Koster rules remain intact even for the
two-electron exchange integrals involving the error
function.\footnote{The mentioned symmetry rules hold due the fact that
  the reference density matrix $P_{\mu\nu}^0$ is diagonal. This can be
  shown by combining (i) the results of Harris \cite{Harris2002}, who
  has given analytical formulas for general two-center two-electron
  integrals of Slater type orbitals with the Coulomb kernel, (ii) the
  series expansion of the error function and (iii) the generalized von
  Neumann expansion for $r_{12}^k$ with $k\ge -1$ given by
  Budzi{\'n}ski and Prajsnar \cite{Budzinski1994}. The applicability
  of the transformation rules follows from the fact that the elements
  of $P_{\mu\nu}^0$ are equal for different magnetic quantum numbers.}
This facilitates the implementation of the present scheme as all
major routines for the Hamiltonian setup can be used without
changes. The integrals itself may be evaluated numerically or
analytically in reciprocal space using the known Fourier transforms
over products of Slater type orbitals \cite{NiehausLR2008} and the
simple transform of the $\text{erf}(\omega r)/r$ kernel (see appendix).

\subsection{The first order Hamiltonian $H_{\mu\nu}^\text{(1)}$}
\label{firo}
Within the two-center approximation of Eq.~(\ref{DFTBme}) the zero order Hamiltonian 
is treated exactly. In contrast, being dependent on the actual molecular density, the
 first order Hamiltonian in Eq.~(\ref{hmn1}) is subjected to additional simplifications 
to avoid numerical quadratures during the run time of the code. As in the DFTB approach, 
products of basis functions on different centers are expressed in the Mulliken approximation:
\begin{equation}
\label{mull}
\phi_{\mu}(\br) \phi_{\nu}(\br) \approx \frac{1}{2} S_{\mu\nu} \left( |\phi_{\mu}(\br)|^2 + |\phi_{\nu}(\br)|^2 \right).
\end{equation}

Introducing further net atomic Mulliken charges: 
\begin{equation}
\label{mullc}
\Delta q_{A} = \frac{1}{2} \sum_{\mu \in A} \sum_\nu \left( S_{\mu\nu} \Delta P_{\nu\mu} + S_{\nu\mu} \Delta P_{\mu\nu} \right), 
\end{equation}

the first term in Eq.~(\ref{hmn1}) simplifies to
\begin{equation}
\label{fr}
\Delta v^\text{fr}_{\mu\nu} = \frac{1}{2} S_{\mu\nu} \sum_{C} \left( \gamma^\text{fr}_{AC} + \gamma^\text{fr}_{BC} \right)\Delta q_{C};\, \mu \in A, \nu \in B.
\end{equation}

The quantities $\gamma^\text{fr}_{AB}$ in the previous equation are
the following two-center two-electron integrals (not to be confused
with the density matrix $\gamma$):
 \begin{eqnarray}
  \label{gammafr}
  \lefteqn{\gamma^\text{fr}_{AB} =}\\ &&\iint  F_A (\bo)
  \left(\frac{1}{r_{12}} +  f^\txsrc[\rho_0](\bo,\bt)  \right)
 F_B (\bt) \,\drodrt,\nonumber
\end{eqnarray}
with the spherical functions 
\begin{equation}
\label{func}
F_A(\br)= \frac{1}{(l+1)^{2}} \sum_{lm} |\phi_{Alm}(\br)|^2.
\end{equation}

Here the averaging over the basis functions ensures that the integral
approximations respect rotational invariance. Formulas equivalent to
Eqs.~(\ref{fr}) and (\ref{gammafr}) appear also in the original DFTB
method \cite{elstner1998scc}, the only difference being that the kernel
$f^\txsrc$ is replaced by the full kernel $f^\text{xc}$ of a
pure density functional in the former treatment. In
Ref. \cite{elstner1998scc}, an approximation for the corresponding integral
was derived by assuming the special form 
\begin{equation}
  F_A(\br) =  \frac{\tau_A^3}{8\pi}  \exp\left(-\tau_A |{\bf r}-{\bf R_A}|\right)
\end{equation}

and evaluation of the Coulomb integral without the contribution of the 
kernel $f^\text{xc}$. The analytical result related the integral value 
to the decay constant $\tau_A$ and the inter-atomic distance $R_{AB}$. An 
independent calculation of the on-site value $U_A=\gamma_{AA}$ (a measure of the 
chemical hardness of element A) now including $f^\text{xc}$, fixed the free 
parameter $\tau_A$ and provided an estimation of screening effects also for the 
off-site elements $\gamma_{AB}$. It follows, that for the range-separated extension of 
DFTB the functional form for $\gamma_{AB}(U_A,U_B,R_{AB})$ can be directly taken 
over from \cite{elstner1998scc}, if the on-site values are computed according to
 Eq.~(\ref{gammafr}), giving rise to modified $\omega$-dependent parameters $U^\text{fr}_A$.

An analogous treatment for the second term in Eqs.~(\ref{hmn1}), involving the long range exchange kernel, leads to the following form:
\begin{gather}
\Delta v^\text{lr}_{\mu\nu} = -\frac{1}{8} \sum_{\alpha\beta} S_{\mu\alpha} S_{\beta\nu} \Delta P_{\alpha\beta} \left( \gamma^\text{lr}_{\mu\beta} + \gamma^\text{lr}_{\alpha\beta} + \gamma^\text{lr}_{\mu\nu}  + \gamma^\text{lr}_{\alpha\nu} \right)\nonumber\\\label{lr}
\gamma^\text{lr}_{\mu\nu} = \gamma^\text{lr}_{AB}\, ;\, \mu \in A, \nu \in B,
\end{gather}

with the integrals 

\begin{equation}
  \label{gamapp}
  \gamma^\text{lr}_{AB} = \iint  F_A (\bo) \erf
   F_B (\bt) \drodrt.
\end{equation}

\begin{figure}
  % Requires \usepackage{graphicx}
  \includegraphics[scale=0.7]{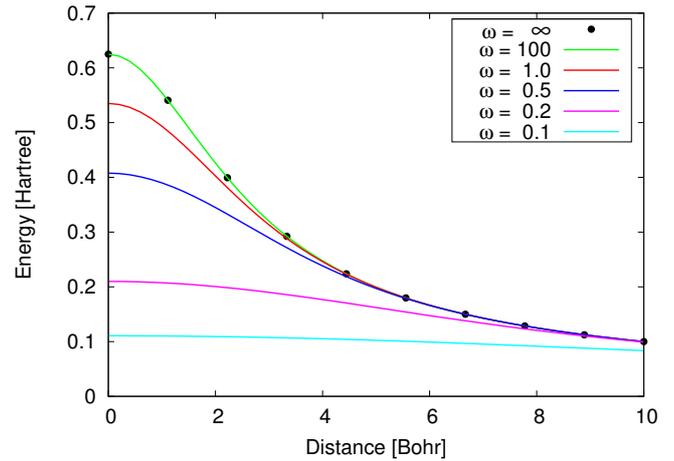}\\
  \caption{The integral $\gamma^\text{lr}_{AB}$ as a function of the inter-atomic distance for different values of the range-separation parameter $\omega$ [bohr$^{-1}$]. The decay constants are $\tau_A=\tau_B= 2.0$ bohr$^{-1}$. The single dots correspond to the function $\gamma$ given in \cite{elstner1998scc}.}\label{gfif}
\end{figure}

As shown in the appendix, this six dimensional integral can be reduced to an efficient one dimensional numerical quadrature:
\begin{equation}
  \label{gamapps}
  \gamma^\text{lr}_{AB} = \frac{2\tau_A^4 \tau_B^4}{\pi R_{AB}} \int_0^\infty  \frac{\sin (q R_{AB})}{q(q^2 + \tau_A^2)^2 (q^2 + \tau_B^2)^2 }  e^{-q^2/4\omega^2} \,dq,
\end{equation}
where the decay constants $\tau_A$ have already been fixed in
the treatment of the $\gamma^\text{fr}_{AB}$ term. Figure (\ref{gfif})
depicts $\gamma^\text{lr}_{AB}$ for various values of the
range-separation parameter $\omega$. In the limit of $\omega$ going to
infinity, the integral needs to  reduce to the mentioned original form
of $\gamma$ given in \cite{elstner1998scc}, because the error function
tends to one in this limit. The results show that this is indeed the
case.

 It should also be mentioned that conventional hybrid functionals
involving a fraction of exact exchange may also be realized in
the present framework. As an example, Eqs.~(\ref{DFTBme},\ref{lr},\ref{gamapp})
with $\omega\to\infty$ give rise to the full Hartree-Fock exchange
potential which might be multiplied with an appropriate constant
factor. Concurrently, the short range exchange in
Eqs.~(\ref{DFTBme},\ref{fr},\ref{gammafr}) has to be replaced by a
conventional full range exchange density functional that is
properly weighted.  

Concerning the Mulliken approximation which leads to matrix elements in Eqs. (\ref{fr},\ref{lr}), it should be 
noted that it completely neglects on-site exchange integrals, which can be restored as describe in Ref. \cite{Niehaus2005a}.
Here it is worth to point out that within the range-separated formalism, these on-site exchange integrals are expected to
 much smaller  (for $\omega$ small) than in the case of HF or pure hybrid functionals.

\section{Total energy and repulsive potential}
\label{reprep}
Considering the results of the previous sections, the total energy of the range-separated DFTB scheme may be summarized as follows:
\begin{equation}
  \label{etotfin}
  E_\ttot = \sum_{\mu\nu} \left ( H_{\mu\nu}^\text{(0)} P_{\mu\nu} 
                       + \frac{1}{2}  H_{\mu\nu}^\text{(1)}\Delta P_{\mu\nu} \right ) + E_\text{rep},
\end{equation}

where we introduced the abbreviation $E_\text{rep}$ for an energy contribution that
 depends solely on the reference density and may hence be precomputed:
\begin{eqnarray}
  \label{erep}
  E_\text{rep} &=& -\frac{1}{2}  \sum_{\mu\nu} \left( 
   v^\th_{\mu\nu}[\rho_0] 
+  v^\txlr_{\mu\nu}[\gamma_0]\right) 
P^0_{\mu\nu} + 
E_\tnn \nonumber\\
&  &+ E_\txsrc[\rho_0]  - \sum_{\mu\nu}  
v^\txsrc_{\mu\nu}[\rho_0]  P^0_{\mu\nu}.
\end{eqnarray}

In the standard DFTB approach the corresponding term is approximated
by a sum of short range pair potentials $E_\text{rep} = \sum_{AB}
V_{AB}(R_{AB})$, which are derived from first principles DFT
calculations \cite{Foulkes1989,elstner1998scc}. A similar approach is possible also in the present context. The additional term in Eq.~(\ref{erep}) stemming from the long range exchange is strictly pairwise and decays as the wave function overlap between two centers. From a practical point of view the direct evaluation of $E_\text{rep}$ might turn out to be more convenient, since each choice of the range-separation parameter $\omega$ necessitates the construction of new pair potentials.

\section{Summary and outlook}
In the last sections we derived the theory for a DFTB formalism with
range-separated exchange-correlation functionals. It is found that the
changes with respect to the original scheme are modest and marginal
code modifications are required for the implementation. In order to
generate the zero order Hamiltonian and on-site two electron integrals,
however, a first principles DFT code featuring range separated
functionals needs to be available. Importantly, the computational cost
of the scheme grows with respect to the original DFTB, but does not reach the formal scaling of $N^4$
for Hartree-Fock based methods. This is due to the integral
approximations applied, which allow to evaluate the terms in
Eq.~(\ref{hmn1}) with cubic scaling. Hence the diagonalization of the
Hamiltonian remains the computational bottleneck.

The presented formalism is also suitable for an straightforward extension of 
the TD-DFTB \cite{Niehaus2001a} approach. The introduction of the long-range term in Eq. \ref{lr} will allow to correctly
describe charge-transfer excitations \cite{tawada2004lrc,baer09,wong09}.

So far we did not discuss the proper choice of the range separation
parameter $\omega$. One of the drawbacks of range separated DFT is the
system dependence of this parameter: different classes of molecules or
even different classes of electronic excitations on the same molecule
may require different values of $\omega$ to achieve accurate
results \cite{Rohrdanz2008}. To our knowledge, a criterion for the
optimal choice of $\omega$, based on first principles is still
lacking. In this regard, the recent work of Baer and coworkers
\cite{Baer2010} provides an elegant expedient. Here $\omega$ is tuned
on a system per system basis to match known conditions for the exact
functional (e.g., the ionization potential being equal to the highest
occupied orbital energy), escaping the need for an empirical fit to a
large training set of molecules. A similar route seems to be promising
for the present approximate scheme. In general, the proposed range
separated DFTB method could be useful to efficiently scan a large
range of parameter values. The optimal value could then be employed in
subsequent first principles calculations.

We hope that the presented theory enlarges the applicability of the
DFTB method to cases where even local or semi-local first principles DFT approaches 
face considerable problems 
The implementation of the formalism is currently under way.

\section{Appendix}
In section (\ref{firo}) the following integral was introduced  
 \begin{equation}
  \label{gamappss}
  \gamma^\text{lr}_{AB} = \iint  F_A (\bo)
 \erf
  F_B (\bt) \drodrt,
\end{equation} 
 where
\begin{equation}
  F_A ({\bf r}) =  \frac{\tau_A^3}{8\pi} \exp\left(-\tau_A |{\bf r}-{\bf R_A}|\right),
\end{equation}
which we attempt to evaluate in Fourier space. Defining
\begin{equation}
  f({\bf q}) = \int f({\bf r}) e^{i{\bf q}{\bf r}} d{\bf r},
\end{equation}

inserting into Eq.~(\ref{gamappss})
twice the identity
  \begin{equation}
  \delta({\bf r}-{\bf r'}) = \frac{1}{(2\pi)^3} \int e^{i({\bf r}-{\bf
      r'}){\bf k}} d{\bf k},
\end{equation}
and integrating out all spatial coordinates,  the remaining integrand is easily seen to factorize into a product of Fourier
transforms
\begin{equation}
  \label{ft}
  \gamma^{lr}_{AB} = \frac{1}{(2\pi)^3} \int  F_A
  (-{\bf q})\left( \int \frac{\text{erf}(\omega
      r)}{r}
  e^{i{\bf q}{\bf r}} d{\bf r}\right) F_B
  ({\bf q}) d{\bf q}.
\end{equation}

The Fourier transforms of the Slater orbitals $F_A$ are known analytically \cite{Belkic1989}
 \begin{equation}
 F_A({\bf q}) = \frac{8\pi\tau_A}{(q^2+ \tau_A^2)^2} e^{i{\bf q} {\bf R}_A},
\end{equation}
and the transform of the long range kernel may be deduced from the transform of the pure $1/r$ kernel given by $4\pi/q^2$ and the integral 

\begin{gather}
 \int_0^\infty \left[ 1 - \text{erf}(a x)\right] \sin b x =
 \frac{1}{b} \left( 1- e^{-\frac{b^2}{4a^2}} \right)\nonumber\\
 \forall a>0,\, b>0,
\end{gather}

given in \cite{Gradsteyn1965} as formula 6.311. We thus obtain

\begin{equation}
 \int \frac{\text{erf}(\omega
      r)}{r}
  e^{i{\bf q}{\bf r}} d{\bf r} = \frac{4\pi}{q^2} e^{-q^2/4\omega^2},
\end{equation}

and after integration over the angular degrees of freedom in Eq.~(\ref{ft}), the end result

\begin{equation}
  \label{gamappfin}
  \gamma^\text{lr}_{AB} = \frac{2\tau_A^4 \tau_B^4}{\pi R_{AB}} \int_0^\infty  \frac{\sin (q R_{AB})}{q(q^2 + \tau_A^2)^2 (q^2 + \tau_B^2)^2 }  e^{-q^2/4\omega^2} \,dq.
\end{equation}

\begin{acknowledgement}
We are especially thankful to Julien Toulouse (UPMC Paris) for fruitful discussions with
respect to this study. T.A.N. would like to thank Thomas Frauenheim for continuous support over the last years. Financial aid by the German Science Foundation (DFG, SPP 1243) is also greatly acknowledged.
\end{acknowledgement}

\bibliographystyle{elsarticle-num}
\bibliography{Combined}
\end{document}